\documentclass[acmtog, nonacm]{acmart}

\usepackage{booktabs} 
\usepackage[ruled]{algorithm2e} 

\SetAlFnt{\small}
\SetAlCapFnt{\small}
\SetAlCapNameFnt{\small}
\SetAlCapHSkip{0pt}

\usepackage{graphicx}
\usepackage{mathtools}
\usepackage{subcaption}
\usepackage{pifont}
\usepackage{multirow}
\usepackage{stfloats}
\usepackage{hyperref}

\renewcommand\footnotetextcopyrightpermission[1]{} 

\renewcommand{\ccsdesc}[2]{}

\newcommand{\figref}[1]{Fig.~\ref{#1}}
\newcommand{\tabref}[1]{Tab.~\ref{#1}}
\newcommand{\eqnref}[1]{Eq.~\ref{#1}}
\newcommand{\secref}[1]{Section~\ref{#1}}

\definecolor{ForestGreen}{HTML}{009B55}

\newcommand{\etal}{\textit{et al.}}
\newcommand{\ie}{\textit{i.e.\xspace}}
\newcommand{\OURS}{ScaffoldAvatar}

\newcommand{\R}{\mathbb{R}}

\begin{document}
\title{\OURS: High-Fidelity Gaussian Avatars with Patch Expressions}

\author{Shivangi Aneja}
\affiliation{
  \institution{Technical University of Munich}
  \city{Munich}
  \country{Germany}}
  \email{shivangi.aneja@tum.de}

\affiliation{
  \institution{DisneyResearch|Studios}
  \city{Zurich}
  \state{CA}
  \country{Switzerland}}
  \email{shivangi.aneja@disneyresearch.com}

\author{Sebastian Weiss}
\affiliation{
  \institution{DisneyResearch|Studios}
  \city{Zurich}
  \state{CA}
  \country{Switzerland}
}\email{sebastian.weiss@disneyresearch.com}

\author{Irene Baeza}
\affiliation{
  \institution{DisneyResearch|Studios}
  \city{Zurich}
  \state{CA}
  \country{Switzerland}
}\email{irene.baeza@disneyresearch.com}

\author{Prashanth Chandran}
\affiliation{
  \institution{DisneyResearch|Studios}
  \city{Zurich}
  \state{CA}
  \country{Switzerland}
}\email{prashanthchandran.pc@gmail.com}

\author{Gaspard Zoss}
\affiliation{
  \institution{DisneyResearch|Studios}
  \city{Zurich}
  \state{CA}
  \country{Switzerland}
}\email{gaspard.zoss@gmail.com}

\author{Matthias Niessner}
\affiliation{%
 \institution{Technical University of Munich}
  \country{Germany}}
  \email{niessner@tum.de}

\author{Derek Bradley}
\affiliation{
  \institution{DisneyResearch|Studios}
  \city{Zurich}
  \state{CA}
  \country{Switzerland}
}  \email{derek.bradley@disneyresearch.com}

\renewcommand{\shortauthors}{Aneja, Weiss, Baeza, Chandran, Zoss, Niessner, Bradley, et al.}

\begin{abstract}
    Generating high-fidelity real-time animated sequences of photorealistic 3D head avatars is important for many graphics applications, including immersive telepresence and movies. This is a challenging problem particularly when rendering digital avatar close-ups for showing character's facial microfeatures and expressions. To capture the expressive, detailed nature of human heads, including skin furrowing and finer-scale facial movements, we propose to couple locally-defined facial expressions with 3D Gaussian splatting to enable creating ultra-high fidelity, expressive and photorealistic 3D head avatars. In contrast to previous works that operate on a global expression space, we condition our avatar's dynamics on patch-based local expression features and synthesize 3D Gaussians at a patch level. In particular, we leverage a patch-based geometric 3D face model to extract patch expressions and learn how to translate these into local dynamic skin appearance and motion by coupling the patches with anchor points of Scaffold-GS, a recent hierarchical scene representation. These anchors are then used to synthesize 3D Gaussians on-the-fly, conditioned by patch-expressions and viewing direction.  We employ color-based densification and progressive training to obtain high-quality results and faster convergence for high resolution 3K training images. By leveraging patch-level expressions, \OURS{} consistently achieves state-of-the-art performance with visually natural motion, while encompassing diverse facial expressions and styles in real time.
\end{abstract}

\begin{teaserfigure}
  \includegraphics[width=1.0\linewidth]{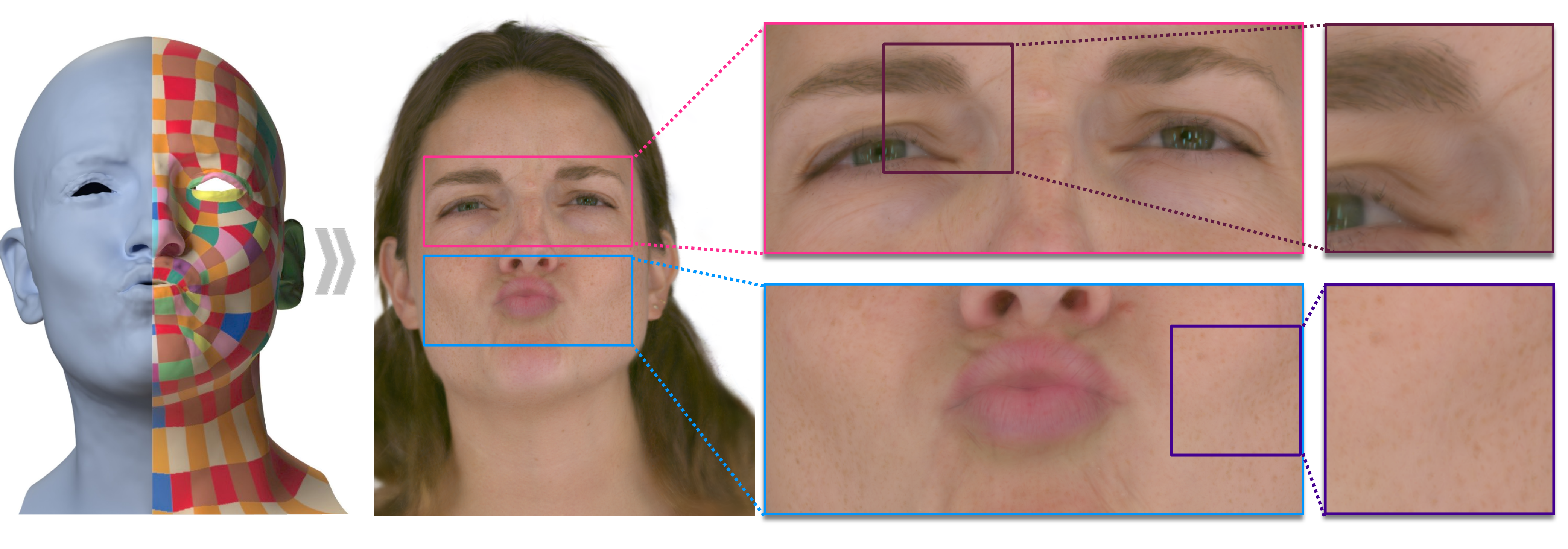}
  \vspace{-1cm}
  \caption
  {Given an input mesh with patch-level localized expressions (left), \OURS{} can synthesize ultra-high fidelity multi-view consistent photorealistic avatars. Our avatars can synthesize high frequency facial details as shown in two-level zoom-ins (right). Our method is capable of generating realistic and high-quality animations including freckles and other fine facial details with real-time rendering.}
    \label{fig:teaser}
\end{teaserfigure}


\maketitle

\pagestyle{empty}

\section{Introduction}
\label{sec:intro}

The creation of realistic digital human avatars is a fast-evolving field with many applications in virtual telepresence, movies and entertainment. The resulting digital avatar should be photorealistic, animatable, allow novel viewpoint rendering in real-time, and must preserve a high degree of spatial fidelity, such as fine-scale expression-specific skin details during animation. Developing a 3D representation that can capture high-frequency appearance and accurate motion dynamics of human heads represents a major challenge, in particular under novel close-up views of the face where the skin exhibits many details.  We require a representation that can jointly reconstruct the facial motion while simultaneously retaining capacity to facilitate high-fidelity re-rendering.

Recent advances in computer vision and graphics, in particular 3D Gaussian Splatting (3DGS)~\cite{kerbl3Dgaussians}, has quickly gained popularity in many application domains due to its simplicity and rendering speed, combined with its differentiable formulation that can be optimized to reconstruct a scene given a set of images from different viewpoints. 
Several works have extended 3DGS to adapt it for multi-vew consistent animatable human avatars~\cite{qian2024gaussianavatars, Xu2023-oq, Giebenhain2024-qk, gaussian_heads}, achieving a new state-of-the-art in realistic digital humans. These existing methods achieve most of the criteria required: photorealism, animatable, novel viewpoints, and real-time, but they fall short on actor fidelity.  For example, the recent GaussianAvatars~\cite{qian2024gaussianavatars} method opened the door to real-time photoreal human heads, but it did not focus on high actor fidelity, so re-rendering appears blurry when zoomed in and expression-dependent appearance details like wrinkles are not supported.  
Neural Parametric Gaussian Avatars (NPGA)~\cite{Giebenhain2024-qk} achieves dynamic expression wrinkles, but still loses fidelity and appear blurry when zoomed in.  Gaussian Head Avatars (GHA)~\cite{Xu2023-oq} make a first attempt at high-fidelity by training on 2K images, which is nearly double the resolution of previous methods.  
However, as we will show in the results, zoom-in regions still lack the necessary details to recover the actor's appearance with sufficient fidelity. 
This lack of fine-scale facial expressions is due to the use of global expression conditioning from existing 3D face models~\cite{giebenhain2024mononphm, gerig2017morphablefacemodels}. 
Such global expression spaces lack expressiveness for fine-grained facial regions, and the inaccurate motion representation due to the low-dimension expression codes leads to blurred out appearance. 
In summary, existing methods lack the representative capacity to re-render photoreal fine scale details in dynamic human faces, which is the problem we tackle in this work.

To this end, we present \OURS, a novel patch-based method for real-time digital avatars that explicitly tackles high spatial fidelity with a novel formulation to increase representative capacity, beyond what is achievable with current state-of-the-art methods. We achieve this by operating on localized patch expressions. 
Our approach has similarities to Scaffold-GS~\cite{Lu2024-av}, a spatially hierarchical 3DGS approach that gives us local control of small facial regions. Our unique architecture provides local network capacity dedicated to recovering Gaussian parameters for the local regions, conditioned on learned local features and local expression parameters.  
The scaffold anchors are attached to patch centers of a tracked 3D mesh, where we use a localized patch-level facial expression, following Chandran~\etal~\shortcite{Chandran2022pbs} who showed that a patch blendshape model is both very expressive and cheap to compute.  The system is trained end-to-end on multi-view dynamic performance sequences with 3K high resolution images. 
Furthermore, unlike existing methods that use positional gradients for densification, we use view-space color gradients to recover sharper skin texture and employ progressive training for faster convergence.

\OURS{} has the capacity to represent fine-scale skin details, including dynamic wrinkles, freckles, and high frequency facial details (as depicted in~\figref{fig:teaser}).  Since our MLPs are conditioned on patch-based expression blendweights, we can regress expression-dependent appearance like wrinkles and blood flow.  Once trained, our method supports real-time inference with control over expression and viewpoint like existing Gaussian-based avatars, but at higher spatial fidelity than current methods. To summarize, we present the following contributions:
\begin{itemize}
    \item A novel method capable of synthesizing ultra-high fidelity photorealistic avatars, due to our unique network formulation similar to Scaffold-GS with conditioning on localized patch-level facial expressions, which can reconstruct finegrained facial details like freckles and wrinkles. 
    \item By leveraging progressive training and view-space color gradient for densification, our method can synthesize sharp skin texture and converge faster for 3K high resolution images.
\end{itemize}

\section{Related Work}
\label{sec:related_work}
We first discuss 3D face and head models that primarily focus on geometry, and then cover more recent photorealistic 3D human head animation techniques.

\subsection{Geometric 3D Face and Head Models}
The seminal work of Blanz and Vetter~\shortcite{3dmm} introduced a model-based approach to represent variations in human faces using PCA. Thereafter, more advanced face models were introduced, including multilinear models for identity and expression~\cite{BolkartWuhrer2015_groupwise, Brunton_ECCV_2014}, as well as recent methods combining linear shape spaces with articulated head parts~\cite{FlameSiggraphAsia2017}. In contrast to these model-based approaches relying on global expression blendweights, techniques based on local patch blendweights~\cite{neumann2013splocs, Chandran2022pbs} can produce much more accurate deformations. For instance, Neumann~\etal~\shortcite{neumann2013splocs} demonstrate that by separating facial performances into localized deformation components, they can produce high-detail deformations like muscle bulges. Similarly, Chandran~\etal ~\shortcite{Chandran2022pbs} show that  operating on small patches on the mesh surface result in more accurately retargeted facial performances. Similar to these works, we leverage local patch blendweights to achieve high frequency details with our patch-based 3DGS avatar.

\subsection{Photorealistic 3D Human Head Animation}

Early approaches for reconstructing photorealistic head avatars were based on NeRFs~\cite{nerf}, which store the radiance field of the scene in a neural network and render novel views of the scene via volume rendering. 
Recently, 3D Gaussian Splatting (3DGS)~\cite{kerbl3Dgaussians} has emerged as an efficient representation for reconstructing fine geometric structures with real-time rendering, leading to a lot of interest in point-based representations for animatable avatars. We discuss both of these approaches below in more detail, in the context of animatable human head avatars.

\subsubsection{NeRF Based Reconstruction}
Gafni~\etal~\shortcite{Gafni_2021_CVPR} introduced one of the first methods conditioning NeRFs on the expression features from monocular videos. Following this,  INSTA~\cite{zielonka2022insta} deforms query points to a canonical space by finding the nearest triangle on a FLAME mesh~\cite{FlameSiggraphAsia2017}, combined with InstantNGP~\cite{mueller2022instant} to achieve fast rendering. NeRFBlendShape~\cite{Gao2022nerfblendshape} models a dynamic scene by blending hash tables with 3DMM parameters. AvatarMAV~\cite{xu2023avatarmav} decouples motion and appearance, blending voxel grids only for the motion field. Nersemble~\cite{kirschstein2023nersemble} uses spatio-temporal NeRFs and learns deformation fields for coarse motion and hash grid encodings for fine scale deformations and is trained using high quality multi-view video data.

\subsubsection{3DGS Based Reconstruction}
Zheng~\etal~\shortcite{Zheng2023pointavatar} explored point-based representations with differential point splatting by
defining a point set in canonical space and learning a deformation field conditioned on FLAME’s expression space for animating the avatar. 
GaussianAvatars~\cite{qian2024gaussianavatars} is the seminal work proposing a method for dynamic 3D representation of human heads based on 3DGS by rigging the anisotropic 3D Gaussians to the faces of the FLAME mesh.
GaussianHeads~\cite{gaussian_heads} predict RGB color and opacity in the UV space of FLAME mesh and deform the canonical gaussians using MLP. 
GaussianHeadAvatars (GHA)~\cite{Xu2023-oq} optimizes neural 3D Gaussians with MLP-based deformations driven by the global expression space of BFM~\cite{gerig2017morphablefacemodels}  for handling the dynamics. 
Very recently, Neural Parametric Gaussian Avatars (NPGA)~\cite{Giebenhain2024-qk} also learns canonical 3D Gaussians and deforms them using a MLP conditioned on the rich expression space of neural parametric head models~\cite{giebenhain2024mononphm}.
While both GHA and NPGA are capable of producing quality animations with expression-dependent deformations and refining the results with a screen-space CNN, they cannot produce fine-scale facial details when zoomed-into different regions of the avatar's face. 

In contrast to these techniques based on global expression features, we leverage local patch level expression blendweights and predict patch level deformations and Gaussian parameters, thus our method is capable of producing finer details for facial zoom-ins.

\section{Preliminaries}
\label{sec:preliminaries}
\label{sec:scaffoldGS}

3DGS~\cite{kerbl3Dgaussians} defines the scene as a collection of Gaussian primitives, each parameterized by position $\mu$, scale $s$, rotation quaternion $\boldsymbol{q}$, opacity $\alpha$ and color $\boldsymbol{c}$.
The scene is then rendered into images with a differentiable rasterizer.
This representation, however, is very expensive in terms of memory to store thousands to millions of primitives. More details can be found in the supplemental material.

To improve scene coverage and avoid redundant Gaussian primitives and redundant parameters, Scaffold-GS~\cite{Lu2024-av} introduces a hierarchy of primitives. 
First, the initial scene is discretized into a sparse voxel grid. 
The center of each voxel $v$ at position $\mathbf{x}_v$ is called an \emph{anchor} and is equipped with a local feature vector $f_v\in\R^F$. 
Each visible anchor at point $\mathbf{x}_v$ spawns $L$ Gaussian primitives with each position $\mu_{l}$ defined as:
\begin{equation}
    \{ \mu_1, \ldots , \mu_{L}  \}=\mathbf{x}_v + \{ \mathcal{O}_1, \ldots , \mathcal{O}_{L} \} \cdot s_{v},
\end{equation}
where $\{\mathcal{O}_1, ..., \mathcal{O}_L \} \in \R^{L \times 3}$ represents the learnable offsets and $s_{v}$ is the scaling factor from the associated anchor. These offsets are learned per anchor.
The remaining 3DGS attributes including opacity $\alpha$, color $\boldsymbol{c}$, rotation $\boldsymbol{q}$, and scale $\boldsymbol{s}$ are predicted by individual but global MLPs $F_\alpha, F_c, F_q, F_s$ conditioned on the learned anchor feature vector $f_v$ and view direction $d_v$.
Note that the parameters of each Gaussian primitive per anchor are predicted at the same time, for example opacity is predicted as:
\begin{equation}
    \{\alpha_1, ..., \alpha_{L} \} = F_\alpha(f_v, d_v) .
\end{equation}
Colors $\{c_l\}$, rotation quaternions $\{q_l\}$ and scales $\{s_l\}$ are predicted in a similar fashion. These spawned Gaussian primitives are then rasterized following standard 3DGS~\cite{kerbl3Dgaussians}.

\section{Data Capture and Preparation}
\label{sec:data}

We begin by describing the data we captured for each actor, and the required pre-processing steps.

\subsection{Capture Setup and Mesh Tracking}
\label{sec:captureSetupAndTracking}

Our main capture setup consists of 9 synchronized video cameras spread out in the hemisphere in front of the actor (8 of which are used for training, one center camera for validation).  For the sake of hardware availability, we use a combination of 12MP and 20MP color cameras, with a combination of 60mm and 85mm lenses, all shooting at 24fps.  We capture between 5-8 performance sequences for each actor, where the sequences contain various localized muscle movements, overall facial workouts, natural expressions, and dialog performances. The multi-view sequences are reconstructed to obtain a topology-consistent tracked mesh of the facial skin and head region using the highly accurate tracking method of Wu~\etal~\shortcite{anyma_tracker}, extended to multi-camera inputs.  

A second capture setup, which consists of 8 static DSLR cameras arranged in 4 stereo pairs, is used to scan 20 blendshapes of each actor~\cite{Beeler2010-jo}, which are used to build the tracking model used on the dynamic performances above, as well as to build a local patch-based model that we describe next.

\begin{figure}[t]
    \begin{center}
    \includegraphics[width=1.0\linewidth]{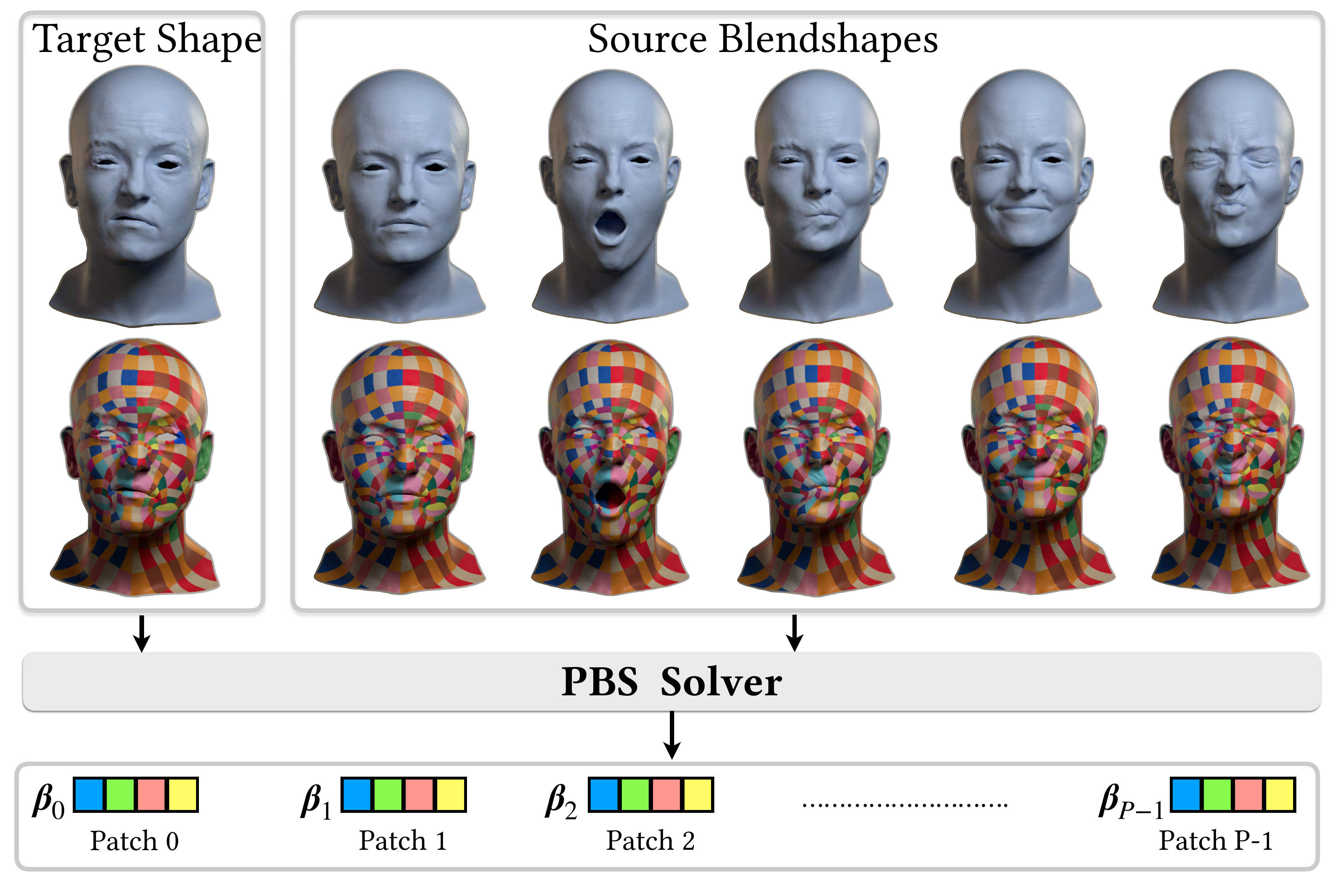}
    \end{center}
    \vspace{-0.5cm}
      \caption{Patch Blendweight Optimziation: Given a target shape from our tracked meshes, we fit patch blendshape model (Eq.~\ref{eq:src_patch_model}) and use our PBS Solver to find optimal per-patch blendweights $ \{ \boldsymbol{\beta}_i \mid i = 0, 1, \dots, P-1 \}$, where $P$ is the total number of patches in our mesh. This process is repeated for every frame in the dataset. The patch layout rendering in the bottom row shows that same patch layout is used for all the shapes.}
    \label{fig:patch_blendshapes}
\end{figure}

\begin{figure*}
    \centering%
    \includegraphics[width=0.95\textwidth]{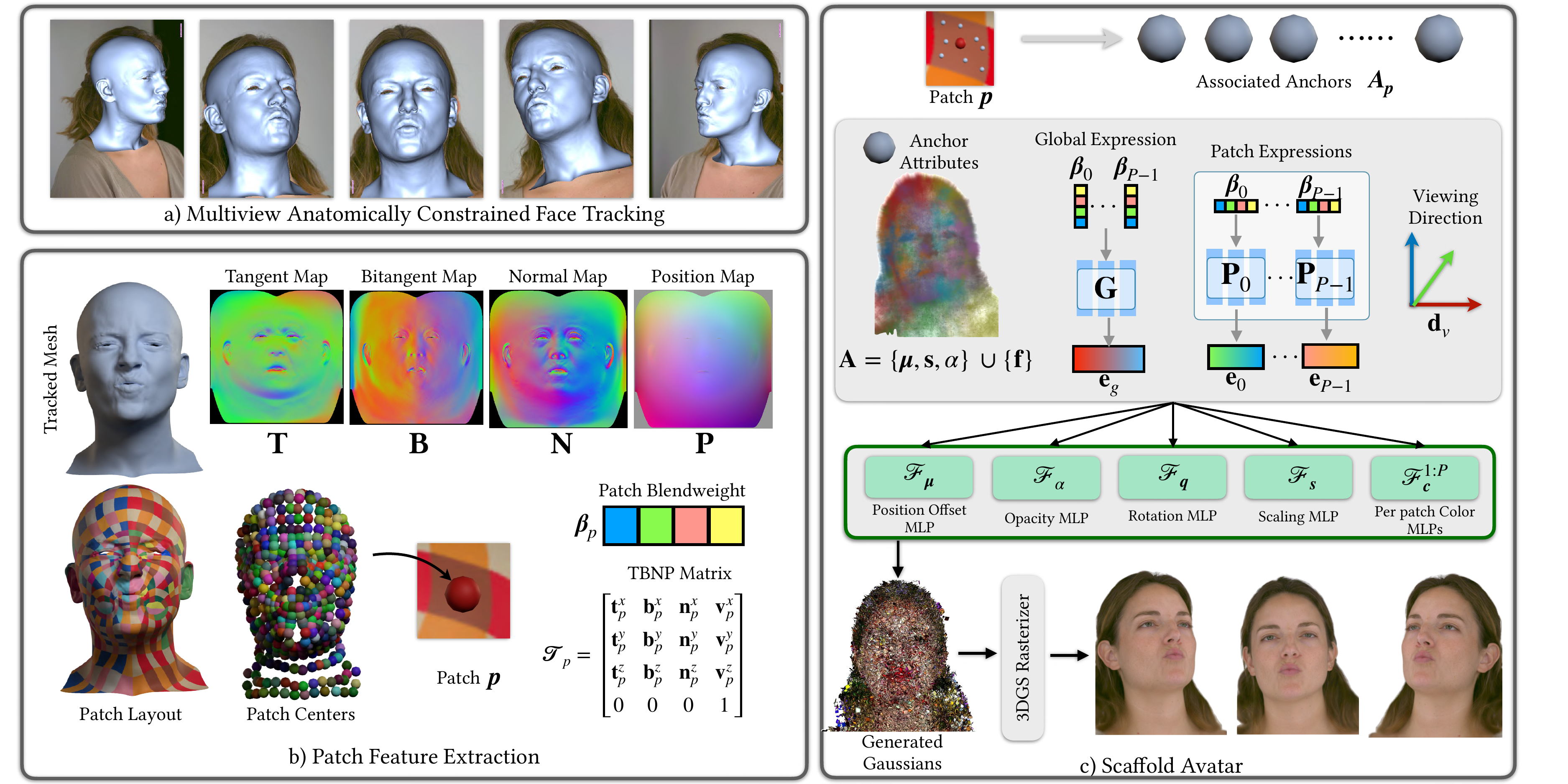}%
    \vspace{-0.4cm}
    \caption{\textbf{Method Overview:} (a) Given a sequence of multiview images, we first run a 3D Face Tracker ~\cite{anyma_tracker} to obtain tracked meshes in consistent topology. (b) Next, we define a patch layout and compute patch centers, together with orientations and positions in the world space (given by a TBNP matrix $\mathcal{T}_p$).  This gives us a per-patch coordinate frame, which we combine with individual patch blendweights $\boldsymbol{\beta}_p$ (obtained from the PBS model Sec.~\ref{sec:patchBlendweightOpt}) for the next steps. (c) Finally, Scaffold-GS anchors $\mathbf{A}_p$ are attached to the patches. Each anchor's attributes, i.e. position $\boldsymbol{\mu}$, scale $\boldsymbol{s}$, opacity ${\alpha}$, and anchor feature $\mathbf{f}$ 
    and optimized together with the global expression MLP $\mathbf{G}$, per-patch expression MLPs $ \{ \mathbf{P}_0 \cdots \mathbf{P}_{P-1} \}$ and scaffold MLPs $ \{ \mathcal{F}_{\boldsymbol{\mu}}, \mathcal{F}_{\alpha}, \mathcal{F}_{\textbf{q}}, \mathcal{F}_{\textbf{s}}, \mathcal{F}_{\textbf{c}}^{1:P} \}$ for decoding the Gaussian features. Note that the Gaussian primitives (position, scale, and orientation) are predicted in the local coordinate frame of the patch and are deformed to global space with the tracked mesh. The result is a high-quality, re-animatable 3DGS-based avatar.}
    \label{fig:pipeline}
\end{figure*}

\subsection{Patch Blendweight Optimization}
\label{sec:patchBlendweightOpt}

We leverage the expressive patch-based facial blendshape model of Chandran~\etal~\shortcite{Chandran2022pbs} to represent dynamic local facial expressions.  To this end, the tracked mesh topology is divided into $P=432$ small overlapping patches as shown in Fig.~\ref{fig:patch_blendshapes}.  For each patch $p$ in the set of all patches $\mathcal{P}$, we build a local blendshape model from the $K=20$ static scans  $\mathcal{S} = \{ \mathbf{S}_i \mid i = 0, 1, \dots, K-1 \}$ as:
\begin{equation}\label{eq:src_patch_model}
    \mathbf{X}_{p}^{\mathcal{S}} = p^{\mathbf{S}_{0}} + \sum_{i=1}^{K-1} \beta_{p,i} \\( p^{\mathbf{S}_i} - p^{\mathbf{S}_0} \\),
\end{equation}
where $\beta_{p,i} \in \mathbb{R}^{K-1}$ are patch blendweights used to linearly blend the patch blendshapes.
Since each patch has its own set of patch blendweights $\beta_{p,i}$, this formulation provides more degrees of freedom for representing the expressions than global blendshape models, \ie, $P \cdot (K-1)=8208$ total expression parameters vs. $100$ of FLAME~\cite{FlameSiggraphAsia2017}.

Next we fit the local blendshape models to all the dynamic captured frames for the corresponding actor.  Specifically, given a captured shape $\mathbf{X}^{t}$ at time $t=1...T$, we solve for all the patch blendweights 
\begin{equation}
    \boldsymbol{\beta}=\{\beta_{p,i}, p=1...P, i=1...K-1\} ,
\end{equation}
such that the resulting deformed patches accurately describe the local skin deformations exhibited in $\mathbf{X}^{t}$, accomplished with a least-squares optimization as:
\begin{equation}
    \mathcal{E}_{ls} = \sum_{p \in \mathcal{P}} \left\| \mathbf{X}_p^{t} - R  \mathbf{X}_{p}^{\mathcal{S}} \right\|^2_2,
\end{equation}
where $\mathbf{X}_{p}^{\mathcal{S}}$ is defined in Eq.~\ref{eq:src_patch_model} and $R$ is global rigid transformation.
The patch coefficients are additionally regularized to remain close to zeros as:
\begin{equation}
    \mathcal{E}_{reg} = \sum_{p \in \mathcal{P}} \sum_{i=1}^{K-1} \| \beta_{p,i} \|_2 ,
\end{equation}
and to stay consistent across adjacent patches,
\begin{equation}
    \mathcal{E}_{o} = \sum_{p \in \mathcal{P}} \sum_{q \in \mathcal{N}(p)}  \sum_{i=1}^{K-1} \| \beta_{p,i} - \beta_{q,i} \|_1 ,
\end{equation}
where $\mathcal{N}(p)$ defines the patches neighboring $p$. The overall loss for fitting the patch blendshape model to a target shape $\mathbf{X}^{t}$ is the weighted sum of these losses as 
\begin{equation}
    \mathcal{E}_{PBS} = \lambda_{ls} \mathcal{E}_{ls} + \lambda_{reg} \mathcal{E}_{reg} + \lambda_{o} \mathcal{E}_{o},
\end{equation}
and final patch blendweights are obtained as:
\begin{equation}
\boldsymbol{\beta}^{\boldsymbol{*}} = \arg\min_{\boldsymbol{\beta}} \mathcal{E}_{PBS}.
\end{equation}
This process is repeated for each of the $T$ shapes in the dynamic data, yielding a sequence of per-frame per-patch expression coefficients $\{\boldsymbol{\beta}^1, \boldsymbol{\beta}^2, ..., \boldsymbol{\beta}^T\}$.  Each actor is processed and trained separately.


\section{Scaffold Avatar}
\label{sec:scaffoldAvatar}

We propose a novel avatar representation that combines the expressive power of local deformation models for human faces with efficient hierarchical scene rendering (similar to Scaffold-GS). Using the local patch blendshape representation (Sec. ~\ref{sec:patchBlendweightOpt}), we assign an anchor to each patch, which spawns multiple Gaussian primitives. These primitives are represented with attributes predicted by MLPs, conditioned on learned anchor features as well as local expression deformation (see \figref{fig:pipeline} for an overview).

A key contribution of our method is how we build the connection between anchors and the patches of our tracked mesh (Sec. ~\ref{method:scf_avatar:patches} and Sec. ~\ref{method:scf_avatar:anchors}) and how to spawn Gaussian primitives around the anchors (Sec. ~\ref{method:scf_avatar:gaussians}). Additional optimization and training details are given in Sec. ~\ref{method:scf_avatar:optimization} and Sec. ~\ref{method:scf_avatar:progressiveTraining}, respectively. Hyperparameters are given in the supplementary document.

\subsection{Defining Patch Centers and Attributes}
\label{method:scf_avatar:patches}
To determine the center of each patch for a given tracked mesh with \( P \) patches, we first calculate the mean position $\mathbf{c}_p$ of all vertices belonging to that particular patch $p$.
Then we designate the closest mesh vertex in Euclidean distance to $\mathbf{c}_p$ as the patch center $\mathbf{v}_p$. 
The patch centers $ \{ \mathbf{v}_p | p=1, \ldots,  P \}$ are determined once and remain fixed throughout the experiments.

To determine the orientation and position of each patch center $\mathbf{v}_p$ in global space, we define the TBNP (Tangent-Bitangent-Normal-Position) matrix $\boldsymbol{\mathcal{T}}_{p}$ as:
\begin{equation}
    \boldsymbol{\mathcal{T}}_{p} = \begin{bsmallmatrix}
\mathbf{t}_p^x & \mathbf{b}_p^x & \mathbf{n}_p^x & \mathbf{v}_p^x  \\[5pt]
\mathbf{t}_p^y & \mathbf{b}_p^y & \mathbf{n}_p^y & \mathbf{v}_p^y \\[5pt]
\mathbf{t}_p^z & \mathbf{b}_p^z & \mathbf{n}_p^z & \mathbf{v}_p^z \\[5pt]
0 & 0 & 0 & 1
\end{bsmallmatrix},
\end{equation}
where $\mathbf{t}_p$, $\mathbf{b}_p$, and $\mathbf{n}_p$ represents the tangent, bitangent and normal vectors at the patch center, and $\mathbf{v}_p$ denotes the position of patch center in global space. These matrices are defined for every patch, computed for every frame, and later used to displace the anchors and Gaussian primitives.

\subsection{Rigging Scaffold Anchors to Patches}
\label{method:scf_avatar:anchors}
For a given patch $p$ we associate a set of anchors in its neighborhood and let these anchors move as the patch moves for different expressions. More specifically, the anchors are static in the local space of its parent patch but change in the global space as the patch moves by transforming them by their parent's patch TBNP matrix.

Let \(\mathcal{A}_p = \{\mathbf{A}_{p,1}, \mathbf{A}_{p,2}, \dots, \mathbf{A}_{p,N}\}\) denote the set of $N$ anchors associated with the patch center \(\mathbf{v}_p\), where each anchor contains the following attributes:
\begin{equation}
    \mathbf{A}_{p,i} = \{ \boldsymbol{\mu}_{p,i}, \mathbf{s}_{p,i}, {\alpha}_{p,i} \} \, \cup  \{ \mathbf{f}_{p,i} \},
\end{equation}
with $\boldsymbol{\mu}_{p,i}$ denoting anchor's position in local space of the patch, $\mathbf{s}_{p,i}$ the anchor's scale, and ${\alpha}_{p,i}$ the anchor's opacity. Similar to Scaffold-GS, we learn per-anchor features $\mathbf{f}_{p,i}$ containing semantic information describing the behaviour of anchor primitives.

Let \(\mathbf{M}_p^{\text{loc}} = \{\boldsymbol{\mu}_{p,1}, \boldsymbol{\mu}_{p,2}, \dots, \boldsymbol{\mu}_{p,N}\}\) denote the set of anchors positions for patch $p$ such that each anchor is represented in the local coordinate space of the patch.
The global positions of the anchors, denoted as \(\mathbf{M}_p^{\text{glob}}\) are given by transforming the anchors as follows:
\begin{equation}
    \mathbf{M}_p^{\text{glob}} = \big\{ \boldsymbol{\mu}_{p,1}^{\text{glob}}, \boldsymbol{\mu}_{p,2}^{\text{glob}}, \dots, \boldsymbol{\mu}_{p,N}^{\text{glob}} \big\}
    \ \text{with} \ 
    \boldsymbol{\mu}_{p,i}^{\text{glob}} = \boldsymbol{\mathcal{T}}_p \cdot 
    \begin{bmatrix}
    \boldsymbol{\mu}_{p,i} \\ 
    1
    \end{bmatrix}.
\end{equation}

Note that the anchors $\mathbf{A}_{p,i}$ do not define a separate rotation $\mathbf{R}$.
Empirically, we found that learning separate rotation for anchors did not help with the quality, hence we omit it and use the rotation of the parent patch $p$ extracted from $\mathcal{T}_p$ as the anchor rotation.

The number of anchors per patch is dynamically optimized during training. New anchors can be added (densification) or old anchors removed (pruning).
Each anchor and their attributes are then used to spawn 3D Gaussian primitives in the global space, which we explain in the next section.

\subsection{Spawning 3D Gaussian Primitives from Blendweights}
\label{method:scf_avatar:gaussians}

One of our core ideas is to leverage per patch expression blendweights $ \{ \boldsymbol{\beta}_1, \dots, \boldsymbol{\beta}_{P} \}$ to synthesizing 3D Gaussian primitives. However, directly using patch blendweights leads to suboptimal results. Therefore, we leverage a set of small per-patch blendweight MLPs \( \mathcal{P}=
\{ \mathbf{P}_1, \dots,  \mathbf{P}_{P}\} \) to map patch-specific blendweights to latent expression features. Each patch expression MLP $\mathbf{P}_i$ processes the blendweights $\boldsymbol{\beta}_{i}$ of the corresponding patch and predicts latent features $\boldsymbol{e}_{p}$ capturing the localized patch expression information as:
\begin{equation}
    \mathbf{e}_p = \mathbf{P}(\boldsymbol{\beta}_p), \quad \forall p \in \{1, 2, \dots, P\}.
\label{eq:expr_latent}
\end{equation}

The per patch blendweight MLPs ensure accurate handling of fine-grained expression details across patches for subsequent processing by Scaffold MLPs. We additionally learn a global expression MLP $\mathbf{G}$ with the concatenated patch blendweights as input for encoding global expression semantics:
\begin{equation}
\mathbf{e}_g = \mathbf{G}(\boldsymbol{\beta}_1, \dots \boldsymbol{\beta}_{P}).
\label{eq:expr_global}
\end{equation}

Similar to Scaffold-GS, we use a fixed number of $L=5$ Gaussian primitives per anchor and use small Scaffold MLPs to learn and compress the 3DGS attributes.
These MLPs directly predict the attributes of all primitives in a single forward pass and are conditioned on anchor attributes,  latent expression features and viewing direction. 
Specifically, we learn a set of Gaussian attribute MLPs $\mathcal{F} = \{ \mathcal{F}_{\boldsymbol{\mu}}, \mathcal{F}_{\mathbf{q}}, \mathcal{F}_{\mathbf{s}}, \mathcal{F}_{\alpha}, \mathcal{F}_{\mathbf{c}}^{1:P} \}$ one for each of the Gaussian attributes, except color where we use one MLP per patch. Empirically we found that this resulted in sharper and more accurate colors. 
Note that in contrast to vanilla Scaffold-GS, we also use an MLP to predict the Gaussian position $\mu$.
We use a sigmoid as the last activation layer for $\mathcal{F}_{\textbf{c}}^{p}$ and $\mathcal{F}_{\alpha}$, and a softplus for $\mathcal{F}_{\textbf{s}}$. We omit patch-anchor association index for ease of notation in the following equations.

Given anchor feature $\mathbf{f}$, patch expression features $\mathbf{e}_{p}$ (\eqnref{eq:expr_latent}), global expression feature $\mathbf{e}_{g}$ (\eqnref{eq:expr_global}), viewing direction $\mathbf{d}_v$, we predict the  attributes for $L$ Gaussian primitives per global MLP as:
\begin{subequations}
\begin{equation}
    \hat{\boldsymbol{\mu}}_{1:L} = \mathcal{F}_{\boldsymbol{\mu}}(\mathbf{f}_{p}; \mathbf{e}_{p}; \mathbf{e}_{g}; \boldsymbol{{\mu}} ), 
\end{equation}
\begin{equation}
    \hat{\textbf{s}}_{1:L} = \mathcal{F}_{\textbf{s}}(\mathbf{f}_{p}; \mathbf{e}_{p}; \mathbf{e}_{g}; \textbf{d}_{v} ),
\end{equation}
\begin{equation}
    {\textbf{q}}_{1:L} = \mathcal{F}_{\textbf{q}}(\mathbf{f}_{p}; \mathbf{e}_{p}; \mathbf{e}_{g}; \textbf{d}_{v} ), 
\end{equation}
\begin{equation}
    \hat{\alpha}_{1:L} = \mathcal{F}_{\alpha}(\mathbf{f}_{p}; \mathbf{e}_{p}; \mathbf{e}_{g}; \textbf{d}_{v} ). 
\end{equation}
\end{subequations}

The final Gaussian colors are predicted by the associated patch MLP $\mathcal{F}_{\textbf{c}}^p$ to which the anchor belongs as:
\begin{equation}
    \hat{\textbf{c}}_{0:L-1} = \mathcal{F}_{\textbf{c}}^{p}(\mathbf{f}_{p}; \mathbf{e}_{p}; \mathbf{e}_{g}; \textbf{d}_{v} ). 
\end{equation}

The final Gaussian scale $\textbf{s}_{g}$ is the anchor scale $\textbf{s}$ combined with Gaussian's local scale $\hat{\textbf{s}}$ as
\begin{equation}
    \textbf{s}_{g} = \hat{\textbf{s}} \; \textbf{s}.
\end{equation}

Given an anchor's position in global space $\boldsymbol{\mu}^{\text{glob}}$, scaling $\textbf{s}$, rotation matrix $\mathbf{R}$ and predicted Gaussian's local position $\hat{\boldsymbol{\mu}}$, the final Gaussian's position in global space is calculated as:
\begin{equation}
    \boldsymbol{\mu}_{g} = \boldsymbol{\mu}^{\text{glob}} + \textbf{s}.\mathbf{R}.\hat{\boldsymbol{\mu}}.
\end{equation}

The Gaussian's rotation quaternion in the global space $\mathbf{q}$ is predicted directly from rotation MLP $\mathcal{F}_{\mathbf{q}}$. Note that during training we only render Gaussians whose parent's anchors are visible from the given viewpoint and have an opacity $\alpha$ above the threshold $\tau$:
\begin{equation}
    \alpha_{g} = (\alpha > \tau) \; \cap \hat{\alpha}.
\end{equation}

The final Gaussian attributes in the global space are passed into the 3DGS rasterizer $\mathcal{R}$ and the RGB image is rendered as:
\vspace{-0.1cm}
\begin{equation}
    I_{\text{RGB}} = \mathcal{R}(\boldsymbol{\mu}_g; \mathbf{s}_g; \mathbf{q}; \alpha_{g}; \mathbf{c}_g).
\end{equation}

\subsection{Optimization and Regularization}
\label{method:scf_avatar:optimization}
Unlike GHA~\cite{Xu2023-oq}, having a fixed number of Gaussian primitives is insufficient for capturing fine facial details. For instance, representing dynamic wrinkles need more Gaussians with dynamically changing color based on expressions compared to a constant mole on the skin. Therefore, we need an adaptive density control strategy that can add and remove anchors based on view-space gradient of the spawned Gaussians. 

We found in our experiments that standard position-based densification used by current methods is slower to optimize and does not lead to sufficient densification to render fine facial features. Instead we use a color-based densification strategy that converges much faster and can more accurately reconstruct fine facial details. We follow the densification strategy similar to Scaffold-GS, except that we use screen space color gradient as our heuristic for densification and pruning anchors.
Since we bind anchors to the patch centers, we additionally track one more parameter with the anchor attributes $\mathcal{A}$, which is the index of its parent patch. This way we also keep track of the anchors attached to each patch and ensure that every patch always has at least one anchor attached.

\begin{figure}[t]
    \begin{center}
    \includegraphics[width=1.0\linewidth]{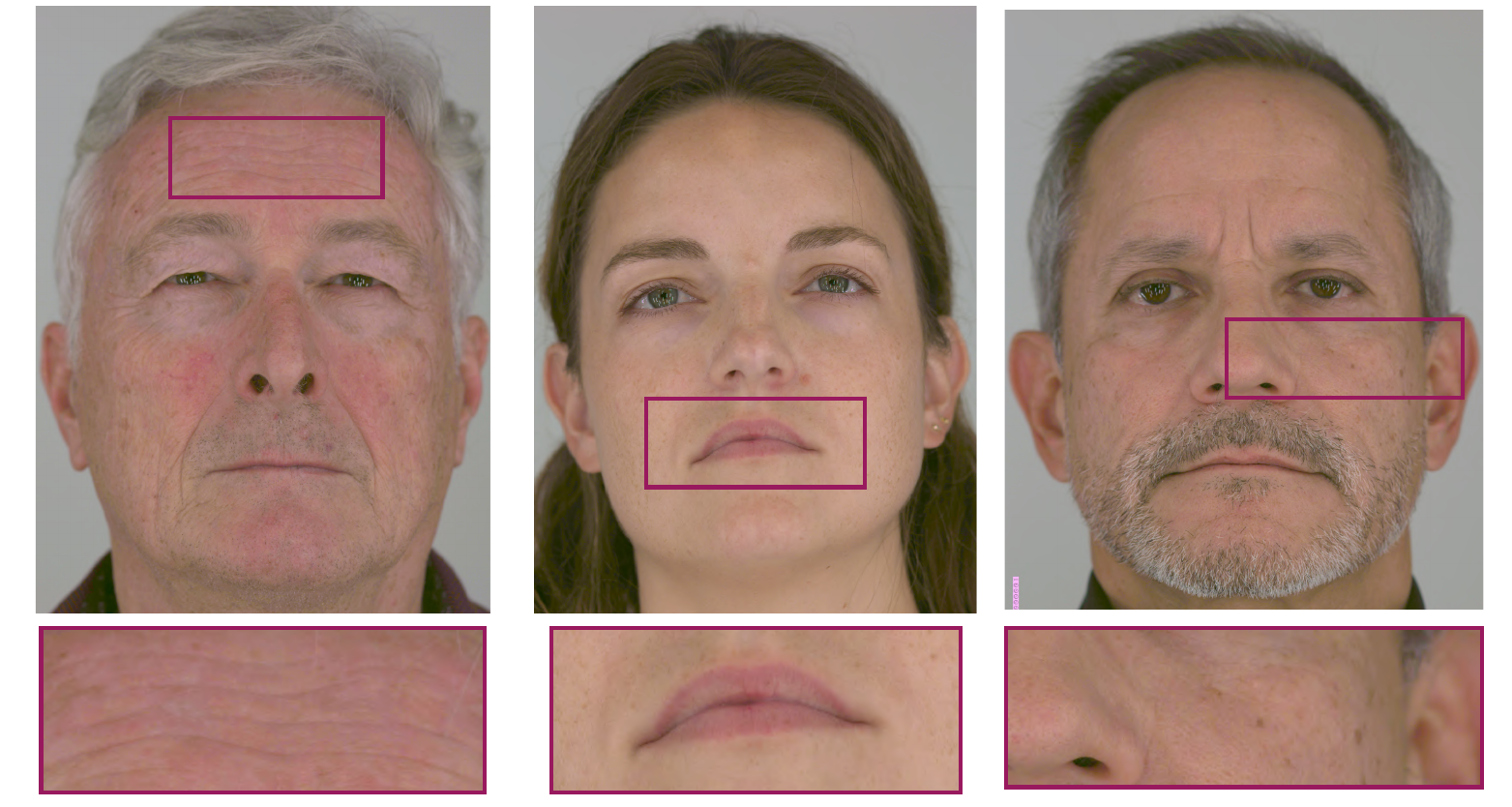}
    \end{center}
    \vspace{-0.4cm}
      \caption{Dataset: One frame selected for each participant from the dataset and corresponding zoom-in for some regions. We selected participants with diverse facial geometry like beard and freckles.}
    \label{fig:dataset}
\end{figure}

We supervise our renderings with L1 and SSIM loss as:
\vspace{-0.1cm}
\begin{equation}
    \mathcal{L}_{\text{rgb}} = (1-\lambda) \mathcal{L}_1 + \lambda \mathcal{L}_\textrm{SSIM}.
\label{eq:loss:rgb}
\end{equation}

To further enhance perceptual fidelity, especially when zooming into different facial regions, we additionally apply the LPIPS~\cite{zhang2018perceptual} loss on local image patches as:
\vspace{-0.1cm}
\begin{equation}
    \mathcal{L}_{\text{patch}} =\frac{1}{J}  \sum_{j=1}^{J} \sum_{k=1}^{K}   \mathcal{L}_{\textrm{LPIPS}}(I_\textrm{RGB}^{j}, I_\textrm{GT}^{j}),
\end{equation}
where   $I_\textrm{RGB}^{j}$ and $I_\textrm{GT}^{j}$ refer to the $j^{th}$ local patch regions from  the rendered and ground-truth multiview images. We use $256 \times 256$ patches  and sample 16 local patches uniformly for the facial area.

Our anchor-patch association ensures that anchors be spatially correlated with their parent patches andstay close, e.g. an anchor for the mouth should not be associated with a patch from the hair.
Similarly, Gaussians should also stay close to their parent anchor. 
To address this and improve locally consistent motion, we apply soft regularization on the anchor and Gaussian positions as:
\vspace{-0.2cm}
\begin{equation}
    \mathcal{L}_{\text{xyz}} = \| \boldsymbol{\mu} \|_2 + \| \hat{\boldsymbol{\mu}} \|_2,
\end{equation}
where $\boldsymbol{\mu}$ and $\hat{\boldsymbol{\mu}}$ refers to anchor and Gaussian position in their respective local space.

We further add a regularization similar to Saito~\etal~\shortcite{saito2024rgca} to ensure that the scale of the Gaussian primitives stays in a reasonable range, defined as:
\begin{equation}
    \mathcal{L}_{\text{scale}} =
\begin{cases}
\frac{1}{\max(s, 10^{-7})} & \text{if } s < 0.1, \\
(s - 10.0)^2 & \text{if } s > 10.0,\\
0 & \text{otherwise.}
\end{cases}
\end{equation}

Our final loss function, therefore can be written as:
\begin{equation}
    \boldsymbol{\mathcal{L}} = \mathcal{L}_{\text{rgb}} +  \lambda_{\text{patch}}\mathcal{L}_{\text{patch}} +   \lambda_{\text{xyz}}\mathcal{L}_{\text{xyz}} +  \lambda_{\text{scale}}\mathcal{L}_{\text{scale}} .
\label{eq:loss_total}
\end{equation}

\begin{figure*}
    \begin{center}
    \includegraphics[width=1.0\linewidth]{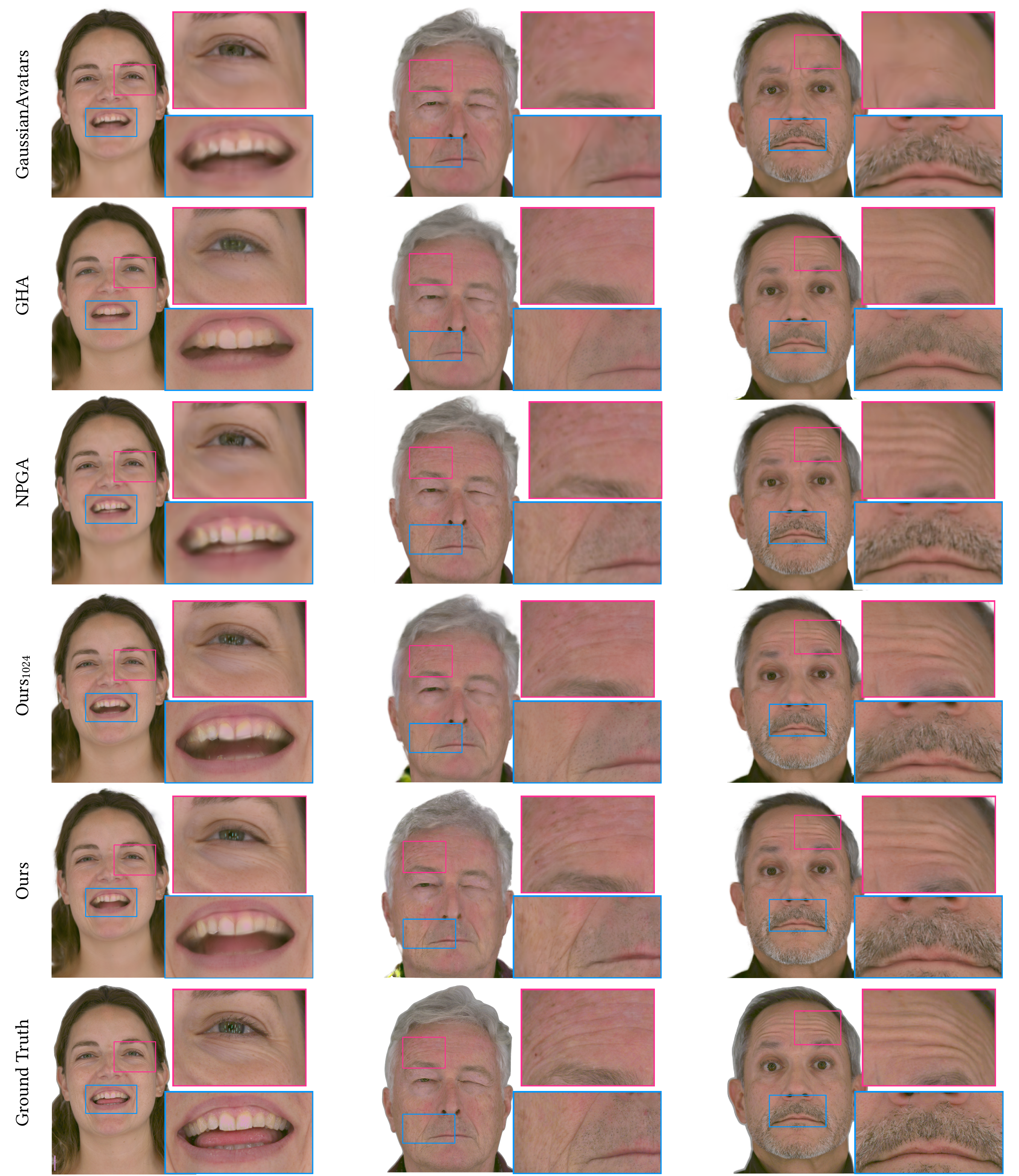}
    \end{center}
    \vspace{-0.4cm}
      \caption{ \textbf{Novel View Synthesis}: Our method trained even at 1K resolution outperforms recent state-of-the-art methods and produces significantly sharper renderings. Refining further with 3K resolution data adds even more details. Our method achieves precise reconstruction like wrinkles around the eyes. We request readers to zoom-into the pdf to see subtle differences. Note that all baseline methods were trained at 1K resolution for consistent comparison.}
    \label{fig:baseline_novel_view}
\end{figure*}

\subsection{Progressive Training}
\label{method:scf_avatar:progressiveTraining}
Since we want to focus on high frequency facial details in different facial regions, it is important to train with high resolution data to get the details upon zooming. However, upon directly training with high resolution images (3K), the method trains too slow and fails to converge in a reasonable time. Therefore, we first train with 1K images and progressively increase the resolution to 3K. This converges much faster while bringing out all the details in different skin areas of the facial region.

\begin{figure*}
    \begin{center}
    \includegraphics[width=1.0\linewidth]{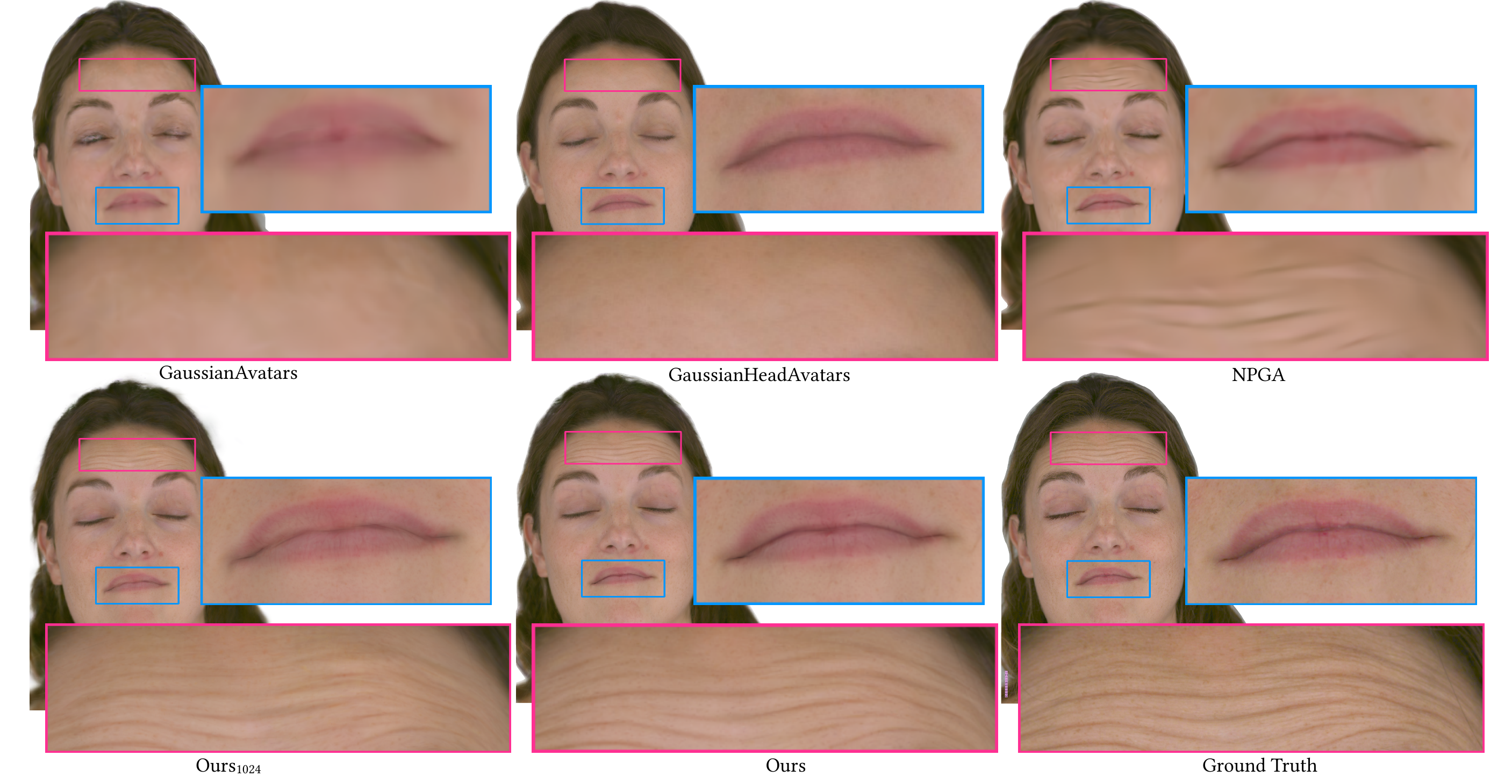}
    \end{center}
    \vspace{-0.4cm}
      \caption{\textbf{Self-Reenactment:} We perform self-reenactment with one held out sequence. Our method achieves ulta-high fidelity results for zoom-ins.}
    \label{fig:baseline_novel_expr}
\end{figure*}

\section{Experiments}

We collected a high-quality dataset captured using the multi-view setup and pre-processing steps described in \secref{sec:data}. We selected three challenging participants with facial details like freckles, wrinkles and facial hair. A snapshot of the dataset is shown in \figref{fig:dataset}. Our model is trained progressively on NVIDIA RTX A6000 (48 GB VRAM) for up to 100,000 iterations (3-4 days) until convergence. On a consumer GPU (NVIDIA RTX 4070Ti 12GB VRAM), it achieves 100.78 FPS for 
$\text{Ours}_\text{1024}$ (our 1K res model) and 76.91 FPS for our full resolution model, making it suitable for interactive applications.

\subsection{Avatar Reconstruction and Animation}
We evaluate \OURS{} on two tasks, (a) Novel View Synthesis: driving the avatar with training sequences them from held-out/novel viewpoints, and (b) Self-Reenactment: animating the avatar with unseen expressions from held-out viewpoints.
For each subject, we downsample images to 3072 $\times$ 2304, and first train at 1024 $\times$ 768 and progressively refine at 3K res. We train with all except one workout sequence (used as held out sequence for self-reenactment) and all views except the front view which is used for novel view synthesis. 

\begin{table}[t]
\caption{\textbf{Quantitative Comparison.} We compare to recent state-of-the-art baseline methods for novel view synthesis (NVS) and Self-Reenactment. }%
    \vspace{-1em}%
    \begin{center}%
    \resizebox{1.0\linewidth}{!}{%
    \begin{tabular}{l ccc | ccc}%
        \toprule
        \multirow{2}{*}{Method} & \multicolumn{3}{c}{NVS} & \multicolumn{3}{c}{Self-Reenactment} \\
        \cmidrule(lr){2-4} \cmidrule(lr){5-7}
         & PSNR $\uparrow$ & SSIM $\uparrow$ & LPIPS $\downarrow$ & PSNR $\uparrow$ & SSIM $\uparrow$ & LPIPS $\downarrow$ \\
        \midrule
        $\text{GA}$~\shortcite{qian2024gaussianavatars} & 26.76 & 0.9082 & 0.1489 & 24.52 & 0.9078 & 0.1952 \\
        $\text{GHA}$~\shortcite{Xu2023-oq} & 28.93 & 0.9366 & 0.1335 & 26.82 & 0.9395 & 0.1911 \\
        $\text{NPGA}$~\shortcite{Giebenhain2024-qk} & 30.15 & 0.9398 & 0.1312 & 27.44 & 0.9335 & 0.1857\\
        $\text{Ours}_\text{1024}$ & 32.19 & 0.9653 & 0.12597 & 29.82 & 0.9515 & 0.1813 \\
        \midrule
        $\text{Ours}$ & \textbf{34.48} & \textbf{0.9712} & \textbf{0.1259} & \textbf{30.37} & \textbf{0.9540} & \textbf{0.1797} \\
        
        \bottomrule
    \end{tabular}}
    \end{center}
    \label{tab:baseline_comp}
    \vspace{-0.5cm}
\end{table}

\subsection{Metrics}
We report the results on a held-out sequence (self-reenactment) and evaluate standard perceptual quality metrics (PSNR, SSIM), and perceptual metric LPIPS~\cite{zhang2018perceptual}. We additionally report results for novel view synthesis (NVS), compare all methods against held-out camera view from the train set. We use BackgroundMattingV2~\cite{BGMv2} to remove the background before training. We train all compared baseline methods at 1K resolution and compare against our method trained at both 1K and 3K resolution. Similar to ~\cite{Giebenhain2024-qk}, we mask out the torso and neck and report metrics for the facial region. 

\vspace{-0.2cm}
\subsection{Results}
\subsubsection{Baseline Comparisons}
We compare against recent state-of-the-art methods for NVS and Self-Reenactment. All baseline methods are trained at 1K resolution using the default hyperparameters proposed in their original implementation. For a fair comparison, we also show results of our method trained at 1K. 
GaussianAvatars~\cite{qian2024gaussianavatars} generates blurry results and cannot synthesize expression dependent details like dynamic wrinkles. GaussianHeadAvatars~\cite{Xu2023-oq} generates grid-like artifacts due to super-resolution CNN. While NPGA~\cite{Giebenhain2024-qk} can synthesize dynamic wrinkles and more facial details, it still produces blurry results and artifacts for zoom-ins. Our method outperforms these baselines even at 1K resolution. Our full model trained at 3K consistently achieves better results over all the baselines on both tasks perceptually as well as quantitatively. Qualitative results for NVS are shown in \figref{fig:baseline_novel_view} and for Self-Reenactment in \figref{fig:baseline_novel_expr}. Quantitative results averaged over all subjects are shown in \tabref{tab:baseline_comp}.

\begin{figure*}[t]
    \begin{center}
    \includegraphics[width=0.95\linewidth]{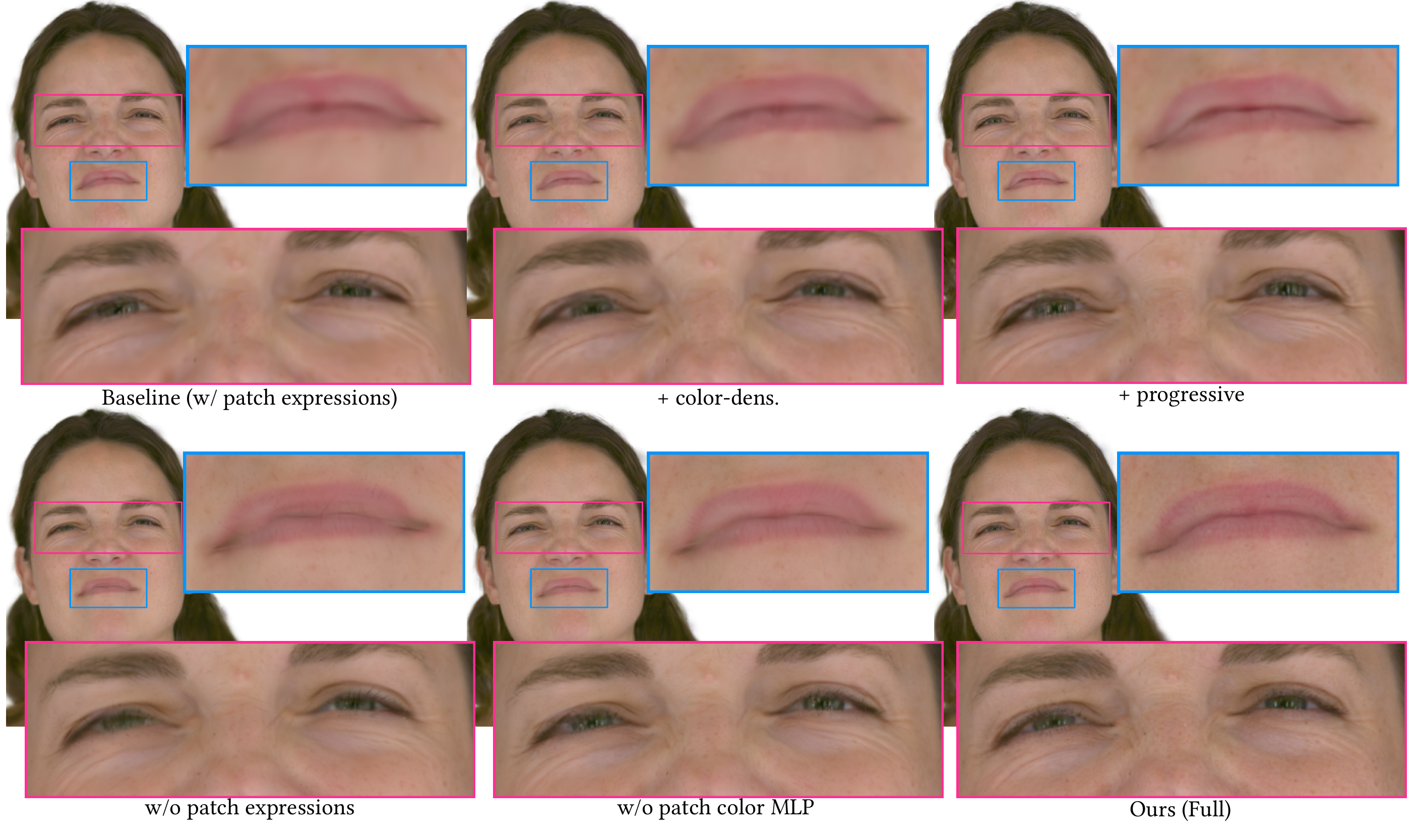}
    \end{center}
    \vspace{-0.6cm}
      \caption{\textbf{Ablation Study:} The na{\"i}ve baseline of our method "Baseline (w/ patch expressions)" trained with position-based densification and without progressive training produces blurry results. Training with color-based densification (+ color-dens.) improves sharpness and is further refined by progressive training (+progressive). Without patch expressions, our method presents artefacts around lip boundaries and produces relatively blurry results. Without using per-patch color MLP, normal zoom-ins  look good but close zoom-ins are blurry. Our full model leveraging all these components achieves the best results. 
      }
    \label{fig:ablation}
\end{figure*}

\begin{table}[t]
\caption{\textbf{Ablation Study.} We perform an ablation study on one of the subjects from our dataset and report results for Novel-View Synthesis (NVS) and Self-Reenactment.}%
    \vspace{-1em}%
    \begin{center}%
    \resizebox{1.0\linewidth}{!}{%
    \begin{tabular}{l ccc | ccc}
        \toprule
        \multirow{2}{*}{Method} & \multicolumn{3}{c}{NVS} & \multicolumn{3}{c}{Self-Reenactment} \\
        \cmidrule(lr){2-4} \cmidrule(lr){5-7}
         & PSNR $\uparrow$ & SSIM $\uparrow$ & LPIPS $\downarrow$ & PSNR $\uparrow$ & SSIM $\uparrow$ & LPIPS $\downarrow$ \\
        \midrule
        $\text{Baseline (w/ patch expressions)}$ & 31.44 & 0.9635 & 0.1772 & 30.34 & 0.9691 & 0.1793 \\
        $\text{+ color-dens}$ & 33.76 & 0.9790 & 0.1615 & 32.63 & 0.9742 & 0.1656 \\
        $\text{+ progressive}$ & 33.53 & 0.9779 & 0.1624 & 33.24 & 0.9775 & 0.1653 \\
        \hline
        $\text{w/o patch expressions}$ & 34.75 & 0.9849 & 0.1291 & 33.66 & 0.9822 & 0.1295 \\
        $\text{w/o patch color MLP}$ & 35.63 & 0.9837 & 0.1075 & 34.39 & 0.9856 & 0.1087 \\
        \midrule
        $\textbf{\text{Ours (Full)}}$ & \textbf{37.03} & \textbf{0.9880} & \textbf{0.0996} & \textbf{35.145} & \textbf{0.9871} & \textbf{0.1065}  \\
        
        \bottomrule
    \end{tabular}}
    \end{center}
    \label{tab:ablation_study}
    \vspace{-0.5cm}
\end{table}

\subsubsection{Ablation Study}
We ablate different design choices of our method on one of the subjects from our dataset. We document these results in \tabref{tab:ablation_study} and show a visual comparison in \figref{fig:ablation}. We started out with a "Baseline" version of our method which serves as the na{\"i}ve baseline of our patch-based avatar. This version is trained with position-based densification, without utilizing progressive training and does not employ per patch color MLP, however it does use per-patch expressions.  While this version of our method can produce light wrinkles, it fails to produce sharp results for fine-scale facial details like freckles on the skin and details on the lips. 

\paragraph{Color-Based Densification.} When changing from view-space positional gradient to view-space color gradient as the heuristic for densifying and pruning the anchors, we notice that our results get sharper. Empirically, we observe that this also converges much faster compared to position based densification.

\paragraph{Progressive Training.} We use color based densification and progressively increase the resolution of the training images. For the first 20,000 iterations we train with 1K images, then for 35,000 iteration we train with 2K and then finally bump up the resolution to 3K. We notice that progressively increasing the image size also improves the quality of our results compared to training directly at 3K.

\paragraph{W/o Patch Expressions.} We use color-based densification and progressive training, but we omit using patch-based expressions and instead concatenate all patch blendweights to obtain a single global expression (Eq.~\ref{eq:expr_global}) which acts as the expression input for our Scaffold MLPs. Note that in this case, we also do not employ per-patch color MLPs but use a single color MLP, since we are operating in global expression space. Although this can produce dynamic wrinkles, we can observe some artefacts around the lips, with intersecting lines between upper and lower lip as shown in \figref{fig:ablation}. 

\paragraph{W/o Patch Color MLP.} Finally, we show the importance of using per-patch color MLPs for predicting view and expression dependent color. Although without using per-patch MLPs, our method already achieves good results, the zoom-ins lack sharpness. Per-patch MLPs can handle super-fine facial details much more effectively, as demonstrated in \figref{fig:ablation} and reflected quantitatively in \tabref{tab:ablation_study}.

\begin{figure*}
    \begin{center}
    \includegraphics[width=1.0\linewidth]{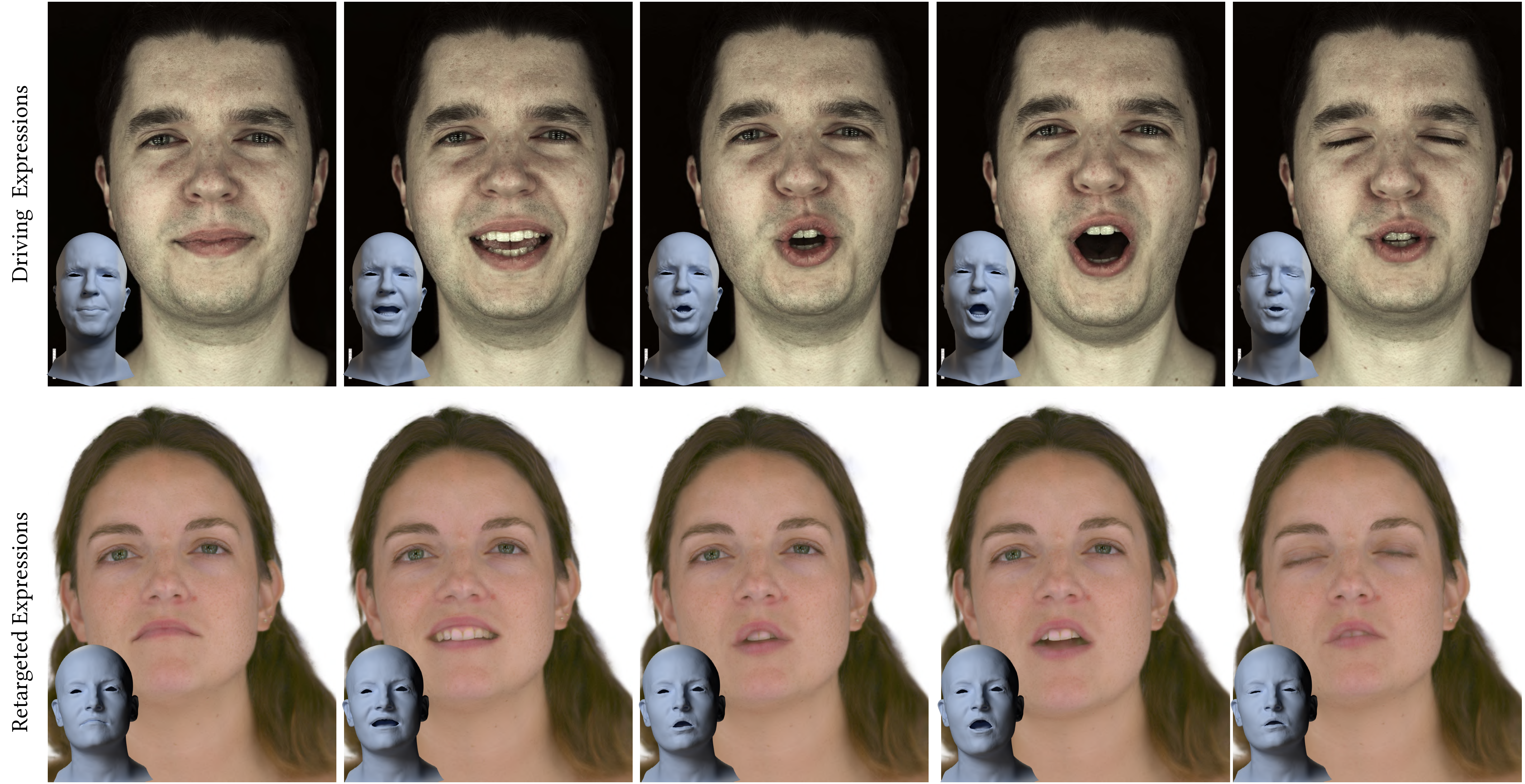}
    \end{center}
    \vspace{-0.4cm}
      \caption{\textbf{Cross-Reenactment}: Qualitative results demonstrating expression transfer from a source actor (top) to our avatar (bottom) across a dialogue sequence. Tracked meshes used in the process are shown in the inset for reference.}
    \label{fig:cross_reenactment}
\end{figure*}

\subsubsection{Cross-Reenactment} We reconstruct the 3D facial performance from the driving actor and perform mesh-based retargeting to obtain the mapped geometry performance of our ScaffoldAvatar (using ~\cite{Chandran2022pbs}). Then, we fit the patch-blendshape model to this 3D performance to obtain local expression blendweights for driving our avatar with the retargeted geometry. To further improve mouth interior, we added 28 additional patches by leveraging tooth geometry. We train our avatar with expression sequences and evaluate retargeting for a dialogue sequence in \figref{fig:cross_reenactment}.

\subsection{Limitations} While our method synthesizes high fidelity avatars, it has some limitations. Our template mesh cannot model accessories like eyewear or headwear, and does not include eyes, so eyeball rotation may not be correctly animated. Since we focus on facial microdetails (wrinkles/freckles), we do not track or reconstruct tongue, so certain expressions with tongue motion might not be correctly tracked.

\vspace{-0.2cm}
\section{Conclusion}
In this work, we propose \OURS, a novel method based on patch blendweights for synthesizing high fidelity 3DGS-based avatars. The key idea of our method is to leverage per-patch blendweights to learn local deformations on the surface in order to produce more accurate and sharper results. We use anchor-based scaffolding and bind them to the patch centers in order to learn high frequency facial details. We utilize effective densification based on color gradient and patch perceptual loss to obtain sharper results. Our method outperforms current state-of-the-art methods, especially when zoomed into the facial regions. We believe that our method takes an important step towards avatar reconstruction and animation with high fidelity facial details.

\begin{acks}
We would like to thank the capture subjects for participating in our dataset. We would also like to thank Lo\"{i}c Ciccone for help with the Patch Blendweight Optimization codebase, Xinya Ji, Yingyan Xu and Christopher Otto for helpful discussions, Pranjal Singh Rajput for proof-reading and Angela Dai for the video voiceover. 
\end{acks}

\clearpage
\bibliographystyle{ACM-Reference-Format}
\bibliography{bibliography}

\newpage

\appendix
\section*{Supplementary Material}
\addcontentsline{toc}{section}{Supplementary Material}


We provide additional details about 3D Gaussian Splatting in Section~\ref{sec:3dgs}, training details in Section ~\ref{sec:train_details}, and further discussion of baseline methods in Section ~\ref{sec:baselines}.

\section{3D Gaussian Splatting}
\label{sec:3dgs}

Recently, 3D Gaussian Splatting~\cite{kerbl3Dgaussians} has emerged as a promising approach to represent a static scene explicitly with anisotropic 3D Gaussians that are optimized given multiview images and camera poses. Specifically, the per Gaussian geometric parameters include a mean position $\boldsymbol{\mu} \in \mathbb{R}^{3} $ and a covariance matrix $\Sigma \in \mathbb{R}^{3 \times 3}$ as:

\begin{equation}
    G(\boldsymbol{x}) = e^{- \frac{1}{2} (\boldsymbol{x}-\boldsymbol{\mu})^T \Sigma^{-1} (\boldsymbol{x}-\boldsymbol{\mu})   }.
\end{equation}

To ensure that the covariance matrix $\Sigma$ is positive semidefinite during gradient-based optimization, the Gaussian is represented as a 3D ellipsoid with scaling matrix $S$ and rotation matrix $R$ as:
\begin{equation}
    \Sigma = RSS^{T}R^{T},
\end{equation}
where scale is represented using a scaling vector $\boldsymbol{s} \in \mathbb{R}^3$ and a quaternion $\boldsymbol{q} \in \mathbb{R}^4$ is used for rotation. 

The appearance of a Gaussian is represented by an opacity $\alpha$ and a view-dependent color $\boldsymbol{c}_i$ parameterized by Spherical Harmonics (SH)~\cite{sph_irradiance}, which are optimized together with the geometric parameters.  To render an image, the 3D Gaussians are splatted to screen space.  Each pixel color $\boldsymbol{C}$ is computed by blending all $N$ Gaussians overlapping the pixel as:

\begin{equation}
    \boldsymbol{C} = \sum_{i=1}^{N} \boldsymbol{c}_i \alpha_i \prod_{j=1}^{i-1} (1-\alpha_j).
\end{equation}

A differentiable tile rasterizer enables real-time rendering. To handle complex scenes and respect visibility order, depth-based sorting is applied to the Gaussian splats before blending.

\section{Network and Training Details} 
\label{sec:train_details}

\paragraph*{Network Architectures}
\OURS{} uses several MLPs to predict expression features and Gaussian primitives.
First, the patch expression MLP $\mathbf{P}_i$ (Eq. (13) in the main document) uses an MLP with $K-1=19$ input dimensions, one hidden layer with 32 channels and ReLU activation, and predicts a 32-dimensional patch latent feature $\mathbf{e}_{p,i}$.
The global expression MLP $\mathbf{G}$ (Eq. (14) in the main document) concatenates all input blendshapes to a 8208-dimensional input, uses one hidden layer with 32 channels and ReLU activation, and predicts a 32-dimensional global expression latent feature $\mathbf{e}_g$.

Per anchor $i$ we learn a 32-dimensional anchor feature vector $\mathbf{f}_{p,i}$.
The Gaussian attribute MLPs (Eq. (15) in the main document) are then defined as following:
The offset MLP $\mathcal{F}_{\boldsymbol{\mu}}$ takes as input the anchor features $\mathbf{f}_{p,i}$, the global expression feature $\mathbf{e}_g$, the local expression feature $\mathbf{e}_{p,i}$ and the 3D-position of the current anchor and concatenates them to a 99-dimensional input. Then one hidden layer with 32 channels and ReLU activation, followed by one output layer predicting the $3L=15$ output offsets (xyz for the 5 Gaussian primitives per anchor).
The scale MLP $\mathcal{F}_{\textbf{s}}$, rotation MLP $\mathcal{F}_{\textbf{q}}$, and opacity $\mathcal{F}_{\alpha}$, and color MLP $\mathcal{F}_{\textbf{c}}^p$ are defined similarly, with the exception that the anchor position is replaced by the 3D view direction to model view-dependent effects.
Note that $\mathcal{F}_{\textbf{s}}$, $\mathcal{F}_{\textbf{q}}$, and $\mathcal{F}_{\alpha}$ are defined globally, but color $\mathcal{F}_{\textbf{c}}^p$ is defined per patch with different weights per patch.

For faster network inference and training, we don't instantiate the per-patch MLPs ($\mathbf{P}_i$ and $\mathcal{F}_{\textbf{c}}^p$) $P=432$ times, which would lead to a very large graph during backpropagation.
Instead, we each type of MLP at the same time across all patches using batched matrix multiplications.

\paragraph*{Training Settings and Loss Parameters}
\OURS{} is implemented in PyTorch, we use Adam optimizer for parameter optimization. We use the same set of hyperparameter values across different subjects. We start densification after 2000 iteration and apply densification/pruning after every 500 iterations. We start training our method at 1K resolution with learning rates and exponential decay similar to Scaffold-GS~\cite{Lu2024-av} for 50,000 iterations. For progressive refinement at 3K we use the learning rates to which training at 1K converged and train with Adam optimizer and do not use any more decay and train for 100,000 iterations.  For loss terms, we use $\lambda=0.2$ in Eq.21, $\lambda_{\text{xyz}}=0.001$, $\lambda_{\text{patch}}=\lambda_{\text{scale}}=0.01$ in Eq.25. We train with batch size of 1 using 3DGS rasterizer similar to~\cite{kerbl3Dgaussians}. For color-gradient based densification, we use a threshold of 2e-6. We train our method with a batch size of 1 for 100,000 iterations on a single Nvidia RTX A6000 GPU .

For the learning rates, we use the following values as initialization and then decay them exponentially during optimization. We use $lr:\mathbf{x}$ to denote the learning rate of training parameter $\mathbf{x}$.
\newline
Expression MLPs:
\begin{equation}
    \begin{aligned}
        lr:\mathbf{P}_i&=0.0001 & lr:\mathbf{G}&=0.0001
    \end{aligned}
\end{equation}
Anchors parameters:
\begin{equation}
    \begin{aligned}
        lr:\boldsymbol{\mu}_{p,i}&=0.00016 & lr:\mathbf{s}_{p,i}&=0.005 \\
        lr:{\alpha}_{p,i}&=0.05 & lr:\mathbf{f}_{p,i}&=0.0025
    \end{aligned}
\end{equation}
Gaussian property MLPs:
\begin{equation}
    \begin{aligned}
        lr:\mathcal{F}_{\boldsymbol{\mu}}&=0.01 & lr:\mathcal{F}_{\alpha}&=0.002 \\
        lr:\mathcal{F}_{\textbf{q}}&=0.004 & lr:\mathcal{F}_{\textbf{s}}&=0.004 \\
        lr:\mathcal{F}_{\textbf{c}}^p&= 0.006 &&
    \end{aligned}
\end{equation}

\section{Baselines}
\label{sec:baselines}
We compare against three most recent state-of-the-art methods in the main paper which we explain below.

\paragraph*{GaussianAvatars 
(GA)~\cite{qian2024gaussianavatars}}
GaussianAvatars proposes a method for dynamic 3D representation of human heads based on 3DGS by rigging the anisotropic 3D Gaussians to the faces of a 3D morphable face model. Specifically, the method uses FLAME~\cite{FlameSiggraphAsia2017} as 3DMM due to its flexibility and compactness, consisting of only 5023 vertices. To better represent mouth interior, it generated additional 120 vertices for teeth. Given a FLAME mesh, the idea is to first initialize a 3D Gaussian at the center of each triangle of the FLAME mesh and allow the 3D Gaussians  to move with the faces of the FLAME mesh.

\paragraph*{Gaussian Head Avatars (GHA)~\cite{Xu2023-oq}} GHA is more recent 3DGS based method that learns deformation fields using deep neural network conditioned on tracked expression codes from BFM~\cite{gerig2017morphablefacemodels}. The method uses fixed number of Gaussians and does not performs any densification or pruning during training. However, it start with a very dense set of Gaussians for providing sufficient coverage of expressions. GHA applies super-resolution and screen-space refinement with CNN to further refine the results.

\paragraph*{Neural Parametric Gaussian Avatars (NPGA)~\cite{Giebenhain2024-qk}} NPGA is the most recent and state-of-the-art avatar reconstruction method based on 3DGS. Similar to GHA, NPGA learns deformation by MLPs and applies screen-space refinement, however the method does not perform super-resolution and is conditioned on tracked expression codes from MonoNPHM~\cite{giebenhain2024mononphm}. Since this tracking is purely based on geometric constraints between MonoNPHM's predicted surface and COLMAP point cloud and requires identity codes to be present in MonoNPHM's latent space while optimizing only for expression code, this was infeasible to do on our dataset. Thus, we resort to BFM tracking, which as proposed in their original paper also achieves state-of-the-art results.


\end{document}